\documentclass[12pt,a4paper,final]{article}  %рисунки печатаются

\usepackage{epsfig}
\newcommand{\calH}{{\cal H}}
\newcommand{\ket}[1]{{|#1\rangle}}
\newtheorem{theorem}{Theorem}
\newtheorem{remark}[theorem]{Remark}

\begin{document}

\title{Postcorrection and mathematical model of life in Extended Everett's Concept}
%\title{Postcorrection as a mathematical model of life in Extended Everett's Concept}
%\title{Postcorrection as a mathematical model for providing life in Extended Everett's Concept}
%\title{Mathematical model for theory of consciousness in the framework of Extended Everett's Concept}
\author{Michael B. Mensky
\\P.N. Lebedev Physical Institute, Russian Academy of Sciences\\
53 Leninsky prosp., 119991 Moscow, Russia}
\date{August 20, 2007}

\maketitle

%\newpage

\begin{abstract}
Extended Everett's Concept (EEC) recently developed by the author to explain the phenomenon of consciousness is considered. A mathematical model is proposed for the principal feature of consciousness assumed in EEC, namely its ability (in the state of sleep, trance or meditation, when the explicit consciousness is disabled) to obtain information from all alternative classical realities (Everett's worlds) and select the favorable realities. To represent this ability, a mathematical operation called postcorrection is introduced, which corrects the present state to guarantee certain characteristics of the future state. Evolution of living matter is thus determined by goals (first of all by the goal of survival) as well as by causes. The resulting theory, in a way symmetrical in time direction, follows from a sort of antropic principle. Possible criteria for postcorrection and corresponding phenomena in the sphere of life are classified. Both individual and collective criteria of survival are considered as well as the criteria providing certain quality of life and those which are irrelevant to the life quality. The phenomena of free will and direct sighting of truth (e.g. scientific insight) are explained in these terms. The problem of artificial intellect and the role of brain look differently in the framework of this theory. Automats may perform intellectual operations, but not postcorrection, therefore artificial intellect but not an artificial life can be created. The brain serves as an interface between the body and consciousness, but the most profound level of consciousness is not a function of brain.
\end{abstract}


\newpage

%\tableofcontents

\newpage
\section{Introduction}
\label{Intro}

From the time of creation of quantum mechanics up to now conceptual problems of this theory, or quantum paradoxes, are not solved. They are often formulated as \emph{the problem of measurement}. Various \emph{interpretations of quantum mechanics} are nothing else than attempts to solve this problem. The origin of the measurement problem is the fact that, contrary to classical physics, consciousness of an observer plays an important role in quantum mechanics (this difference may be formulated as the difference between classical and quantum concepts of reality). This allowed for the author to suggest \emph{the theory of consciousness} called Extended Everett's Concept (EEC) starting from the principal points of quantum mechanics. Here we shall introduce a mathematical model for this theory and discuss some principal issues resulting from it. 

The reasoning applied in EEC is following (see Sect.~\ref{sec:2EEC} for detail): 

1)~Commonly accepted \emph{Copenhagen interpretation} of quantum mechanics includes \emph{the reduction postulate} declaring that a quantum system's state is converted, after a measurement, into one of the alternative states corresponding to the alternative measurement outputs (readouts). This postulate contradicts to linearity of quantum mechanics: the state of the measuring device and the measured system should, in linear theory, include all the alternatives as the components of the superposition. In the interpretation suggested by Hew Everett \cite{Everett1957,DeWittGrah73everett} the linearity was taken as a basic principle and therefore \emph{all alternatives were assumed to coexist} (to be equally real). To explain, why any real observer always watches only a single alternative, it was assumed that ``many classical worlds'' (corresponding to various alternatives) exist or, equivalently, that \emph{the observer's consciousness separates the alternatives} from each other (subjectively the observer, when watching some alternative, cannot watch the others). 

2)In the Extended Everett's Concept (EEC) proposed by the author \cite{EEC2000eng,EEC2005eng,EEC2005bkEng,EEC5}, \emph{the observer's explicit consciousness is identified with separating alternatives}. This simplifies the logical structure of the theory and results in new consequences: when the explicit consciousness is disabled (in the states similar to sleep, trance or meditation) one acquires a sort of \emph{``superconsciousness''} being able to take information from all alternatives, compare them with each other and choose the favorable one. This allows one to explain the well known phenomena of free will, absolute necessity of sleep, as well as such unusual phenomena as direct sighting the truth (e.g. scientific insights) and even ``control of reality'' in the form of ``probabilistic miracles''. 

According to EEC, the principal feature of consciousness (of human and, more generally, of any living being) is its ability, overcoming the separation of the alternatives, to follow each of them up to the distant time moment in the future, find what alternatives provide survival and choose these alternatives excluding the rest. The evolution of living matter is thus determined not only by causes, but also by the goals, first of all by the goals of survival and improvement of the quality of life. 
%%%%%%%%%%%%%%%%%%%%%%%%%%%%%%%%%%%%%%%%%%%%%%%%%%%%%
%%%%%%%%%%%%%%%%%%%%%%%%%%%%%%%%%%%%%%%%%%%%%%%%%%%%%

In the present paper we shall introduce the mathematical formalism describing this principal feature of living matter (of its consciousness): the ability to correct its state making use of the information (about the efficient way of survival) obtained from the future. It will be assumed that the evolution of living matter includes the correction providing survival at distant time moments. This correction leaves in the sphere of life only those scenarios of evolution which are favorable for life. Unfavorable scenarios do not disappear from the (quantum) reality but are left outside the sphere of life (are absent in the picture appearing in the consciousness).

This correction (selection of favorable scenarios) is represented by the special mathematical operation which is called \emph{postcorrection}. It corrects the present state of the system in such a way that its future state satisfies a certain criterion. 

After defining the operation of postcorrection, various criteria for postcorrection are considered as well as the corresponding aspects of the phenomenon of life. In particular, a simple mathematical model of a collective criterion of survival is proposed, and the important role played by collective criteria shortly discussed. Stronger criteria providing not only survival but also certain levels of the quality of life, are discussed. It is argued that the postcorrection is possible also according to such criteria which are insignificant for life. Such phenomena as free will and direct sighting of truth may be explained by the action of postcorrection performed according to such criteria. 

The paper is organized in the following way. After a short sketch of EEC given in Sect.~\ref{sec:2EEC}, the operation of postcorrection is defined and the simplest but most important criterion of survival considered in Sect.~\ref{sec:Postcorrection}. In Sect.~\ref{sec:CollectiveLife} a simple example of the collective criterion of survival is given. Various criteria for postcorrection, their classification and the corresponding aspects of the phenomenon of life are discussed. At last, Sect.~\ref{sec:Conclusion} supplies comments on the whole theory. In particular, deep analogy of the postcorrection (providing survival) with the antropic principle is analyzed. 
%%%%%%%%%%%%%%%%%%%%%%%%%%%%%%%%%%
%%%%%%%%%%%%%%%%%%%%%%%%%%%%%%%%%%

\newpage
\section{Extended Everett's Concept (EEC)}
\label{sec:2EEC}

The ``many-worlds'' interpretation of quantum mechanics proposed by Everett in 1957 \cite{Everett1957} has as its starting point linearity of quantum mechanics. The von Neumann's reduction postulate is rejected in this interpretation, and therefore all components of the superposition which correspond to the alternative outputs of a measurement are presented in the measured system's and measuring device's state after the measurement (the only change of the state is entanglement of the measured system with the measuring device). This suggests that the classical alternatives corresponding to various measurement outputs coexist, even though they are conventionally considered to be inconsistent (alternative). 

\begin{remark}
The conclusion about coexistence of various alternatives is made in the Everett's concept in the course of analysis of the procedure of a quantum measurement. In order to go over to EEC, this conclusion has to be considered in a more general context. It is not important that the alternatives may appear as a result of a measurement. The only essential issue is that the state of our (quantum) world may have the form of a superposition, the components of which represent distinct classical pictures. According to Everett, all these ``classical alternatives'' are equally real (coexist). For making the status of these alternative pictures of the world more transparent, they are often called ``Everett's worlds'' \cite{DeWittGrah73everett}, the term ``many-worlds interpretation'' resulting from this. 
\end{remark}

Thus, coexistence of the classical alternatives is predicted by the Everett's concept. However, real observers never see any evidence of this coexistence, always watching only a single alternative. In order to explain this real experience in the framework of the Everett's concept, one has to assume that \emph{the classical alternatives are separated} (disconnected) from each other \emph{in the observer's consciousness}. Then, despite of all alternatives being equally real, an observer, when watching in his (explicit) consciousness one of them, cannot watch at the same time the others. The alternatives coexist but are not ``co-observable''. 

The statement ``the consciousness separates the alternatives'' which is characteristic of the Everett's interpretation has been replaced in the Extended Everett's Concept (EEC) \cite{EEC2000eng,EEC2005eng,EEC2005bkEng,EEC5} with the stronger one: ``the phenomenon of (explicit) consciousness is nothing else than the separation of the alternatives''. Such change of the theory simplifies its logical structure, since two unclear (may be even non-definable) notions are identified with each other and therefore ``explain each other''. These are the notion of ``consciousness'' in psychology and the notion of ``alternatives' separation'' in quantum physics. 

Besides simplifying the logical structure of the theory, this identification results in new very interesting conclusions. In quantum mechanics it becomes clear, in the light of the above identification, why the alternatives are classical (because the classical world is ``locally predictable'' and therefore appropriate for habitation). In psychology it becomes clear why free will is possible and why sleep is absolutely necessary for support of life. Moreover, the strange things characteristic of consciousness, such as direct sighting (revelation) of truth and probabilistic miracles (realization, by the willpower, of events having very low probabilities) may be explained \cite{EEC2005eng,EEC2005bkEng,EEC5}. 

All these conclusions result from the following argument. If the (explicit) consciousness is identical to the separation of the alternatives, then its disappearance (i.e. the transition to unconscious, for example in sleep, trance or meditation) means disappearance (or weakening) of this separation. The consciousness stops to watch the world's state as separated in classical alternatives, but begins to perceive (in some sense or another) this state as a whole. The consciousness stops to watch continuous ``developing'' alternative scenarios, but views instead the reversible evolution of the quantum world i.e. actually four-dimensional image of the world in which all time moments are treated on equal footing. 

In other words, when the explicit consciousness is disabled (in the regime of unconscious), the (implicit) consciousness witnesses, instead of the usual classical world, something quite different, including particularly all classical scenarios in all time moments. Such an image of the world can serve as an enormous ``data base'' allowing particularly comparing various alternative scenarios between each other. This data base may be used first of all for support of life. Indeed, usage of this data base makes possible selecting those scenarios which are favorable for life, i.e. provide survival. Addressing this data base may be performed periodically (for example, in sleep) or even permanently (since many processes in living organisms are regulated unconsciously, with no participation of the explicit consciousness). 

In the next sections we shall suggest a mathematical formalism describing this function of consciousness: its ability to use the information obtained in the future for correcting the present state. To mathematically describe this function, the operation of \emph{postcorrection} will be introduced. This mathematical operation performs such a correction of the state of a ``living system'' which guarantees the required features of its future state. In the simplest case the requirement of survival is meant, but this may also be the requirement of a certain level of quality of life or even the requirement of something that is desirable although not directly connected with the quality of life. 

\newpage
\section{Life as postcorrection with the criterion of survival}
\label{sec:Postcorrection}

Life is a phenomenon which is realized by living matter consisting of living organisms (living beings). Living matter differs from non-living matter in that its dynamics is determined not only by causes, but also by goals i.e. by the state this matter should have in future. First of all the goal of survival (prolongation of life) is important in this context. However, in case of sufficiently perfect forms of life more complicated goals are also actual. They can be formulated in terms of quality of life. 

In the real conditions on Earth, important features of the phenomenon of life are connected with the balance between all organisms. However, the very definition of life and essential features of this phenomenon may be illustrated in case of a single living being. Let us first consider this simple situation (the case of a group of identical living beings will be considered in Sect.~\ref{sec:CollectiveLife}). 

An organism consists of atoms interacting with each other, therefore it is in fact a physical system. According to the modern views this is a quantum system. Let us apply the term ``\emph{living system}'' to refer this quantum system. Denote by ${\cal H}$ a space of states of this system. The state of the environment will be considered (in the simple model we are to discuss) to be fixed.\footnote{This is a sufficiently good approximation if the changes caused by the influence of the living being on its environment is not essential for its life.}

Let $\{L, D\}$ (from the initial letters of the words `life' and `death') be a complete system of orthogonal projectors in the state space ${\cal H}$, so that $L+D=\mathbf{1}$ and $LD=0$. These projectors determine two orthogonal and complementary subspaces $L{\cal H}$ and $D{\cal H}$ in the whole space ${\cal H}$. The subspace $L{\cal H}$ is interpreted as the space of the states in which the body of the living being is acting properly (remains alive). The subspace $D{\cal H}$, vice versa, is interpreted as the space of the states in which the processes of life are seriously violated, the living being is dead. The projector $L$ plays the role of \emph{the criterion of survival}. 

If a quantum system is in the state $\ket{\psi(t_0)}$ at a time moment $t_0$, then its state $\ket{\psi(t)} = U(t,t_0)\ket{\psi(t_0)}$ at time $t$ is determined by the action of the unitary evolution operator $U(t,t_0)$. In case of static environment and invariable properties of the system, the evolution operator depends only on the increment of time: $U(t,t_0)= U_{t-t_0}$. At the moment we shall assume for simplicity this is valid, but the generalization onto the generic situation is straightforward. 

The description of evolution by a unitary evolution operator is characteristic of non-living matter, whose dynamics is determined by causes (by the initial state and Hamiltonian). However, such a description of evolution is not enough for living matter. \emph{The dynamics of a living being is partially determined by goals, i.e. by characteristics of the future state of this living being}. 

In the simplest case the goal is survival. According to this goal the living being has to remain alive, i.e. the state of the living system should be in the subspace $L{\cal H}$ at a distant future moment of time. This is provided by correcting the initial condition in such a way that the evolution of this state brings it into the subspace $L{\cal H}$ in the future. Such correction may be called \emph{postcorrection}. The operation of postcorrection is a correction of the present state of the living system, but it is performed according to the criterion which is applied to the future state of the system. 

Let us consider the simplest example of postcorrection. For simplicity of notation, we shall fix two time moments, ``the present time'' $t=t_0$ and ``the future time'' $t=t_0+T$. Denote by $U_T$ the evolution operator leading from the present time to the future time. 

Let the living system's state at time $t=t_0$ be presented by the vector $\ket{\psi}\in \calH$. If only conventional (characteristic of non-living systems) dynamics act, then after time interval $\tau$ the state vector should be $U_\tau\ket{\psi}$. However, life as a special phenomenon is described only by those scenarios in which the conventional evolution provides survival (prolongation of life). For life prolonging during the time interval $T$, it is sufficient to restrict the initial condition by the requirement for it to be in the subspace $U_T^{-1}LU_T\cdot\calH$. Indeed, any state from this subspace will happen to belong, after the time interval $T$, to the subspace $U_T\cdot U_T^{-1}LU_T\cdot\calH = LU_T\calH = L\calH$, i.e. the living system will remain alive.\footnote{\label{InvarEvol}We took into account that the whole state space $\calH$ is invariant under the unitary evolution, $U_T\calH = \calH$.} 

Thus, the correction selecting the favorable scenarios is described by the projector $L_T = U_T^{-1}LU_T$ which may be called \emph{the postcorrection operator}. The living system's evolution, with the postcorrection taken into account, may be described as a series of short time intervals $\tau$ of the usual (causal) evolutions $U_\tau$, each of them being preceded by the postcorrection $L_T$. This is described as the action of the operator 
\begin{equation}\label{eq:corEvolution}
	U_{n\tau}^{\mathrm{cor}} = \underbrace{U_\tau L_T \cdot \dots \cdot U_\tau L_T \cdot U_\tau L_T}_{n \;\mathrm{times}} 
\end{equation}
which replaces, for the living system, the usual evolution operator $U_{n\tau} = U_\tau  \cdot \dots \cdot U_\tau  \cdot U_\tau$ that had to be taken if the system were non-living.

\begin{remark}
A single period $\tau$ of the evolution according to the equation (\ref{eq:corEvolution}) is represented by the operator $U_\tau^{\mathrm{cor}} = U_\tau L_T = U_{T-\tau}^{-1}LU_T$. Applying operator $LU_T$ to the whole state space $\calH$, we shall obtain $LU_T\calH = L\calH$ i.e. the subspace of alive states (see footnote \ref{InvarEvol}). Therefore, operator $LU_T$ brings any state into an alive state. The operator $U_\tau^{\mathrm{cor}}$ also brings any state into an alive state, $U_\tau^{\mathrm{cor}}\calH \subset L\calH$, provided that $U_{T-\tau}^{-1}L\calH \subset L\calH$. This is a requirement which is necessary for the evolution law (\ref{eq:corEvolution}) being correct. This requirement suggests that the usual causal evolution (represented by a unitary operator and taking into account not only favorable, but all scenarios) cannot convert a dead body into alive one. It is of course evident that living matter has this property. 
\end{remark}
%%%%%%%%%%%%%%%%%%%%%%%%%%%%%%%%%%%%%%%%%%%%%%%%%%%%%
%%%%%%%%%%%%%%%%%%%%%%%%%%%%%%%%%%%%%%%%%%%%%%%%%%%%%

Selecting favorable scenarios does not suggest violating the laws of nature as such. The material world is described as usual by all scenarios obtained by the action of the unitary evolution operators on the arbitrary initial state vectors. This conventional presentation of the evolution of matter is sufficient to describe how non-living matter evolves. However, the phenomenon of life is represented by only a part of the set of all scenarios of evolution. ``Unfavorable'' (for life) scenarios are left ``outside the sphere of life''. The picture appearing in the consciousness of an observer may include only one of the favorable scenarios.\footnote{This expresses the very principle of life, without details like accidents and other casual obstacles for life. In Sect.~\ref{sec:CollectiveLife} we shall consider ``programmed death'' of individuals necessary for life of a group (collective).} Subjectively this looks as if the living being could find out what should be its state in a distant time $t_0+T$ and correct the state at time $t_0$ in such a way that it provides being alive at time $t_0+T$. 

It could be not quite clear what is meant by the words ``the unfavorable scenarios are left outside the sphere of life''. To clarify this, let us reformulate this statement in the language utilized in the preceding works on EEC \cite{EEC2000eng,EEC2005eng,EEC2005bkEng,EEC5} (see also Sect.~\ref{sec:2EEC}), however with the help of the mathematics introduced above. 

In the preceding works the (explicit) consciousness is identified with the separation of the alternatives. In the transition to the regime of unconscious (``at the edge of (explicit) consciousness'') the separation of the alternatives disappears, and the possibility arises for the (implicit) consciousness to compare all alternatives between each other, select favorable ones and discard the rest. How could this be expressed in the language of mathematical formulas? 

Let the set of the (quasiclassical) alternatives at the present time be defined as the set of subspaces $\left\{\calH_i\right\}$. Assume that the favorable (providing survival in the time interval $T$) are the alternatives $i\in I$, while the rest alternatives $i'\in \bar{I}$ (where $I \bigcup\bar{I}$ is the set of all alternatives) are unfavorable. This suggests that $LU_T \calH_i = U_T \calH_i$ for $i\in I$ and $LU_T \calH_{i'} = 0$ for $i'\in \bar{I}$. Therefore, the postcorrection operator $L_T = U_T^{-1}LU_T$ conserves any ``favorable alternative subspace'' and annihilates any unfavorable one, $L_T \calH_i = \calH_i$ for $i\in I$ and $L_T \calH_{i'} = 0$ for $i'\in \bar{I}$.\footnote{In this reasoning we started from the verbal formulation of EEC given earlier. The real situation is very close to this, differing only in that the sets $I$ and $\bar{I}$ do not necessarily cover the set of all alternatives: the alternatives (subspaces) which are intermediate between completely favorable and completely unfavorable may exist.} 

Therefore, ``to stay in the sphere of life'' means ``to leave only favorable (for life) alternatives in the picture appearing in the consciousness''. The rest alternatives (subspaces) do not disappear (this would be the violation of the laws of nature), but simply disappear from the sphere embraced by the consciousness of the living being. 

From this point of view the statement that the phenomenon of life is described by postcorrection performed according to the criterion of survival is in fact not a postulate but only a mathematical form of the definition of life. Any reasonable definition should differ from it only in details, but not in principle. Indeed, the essence of the phenomenon of life reduces to a strategy of survival, and the efficient survival is provided only by estimating the future of a living system (from the point of view of its survival) and by the corresponding correction of the system's present state. 

Some remarks should be made about the evolution law (\ref{eq:corEvolution}). 
\begin{remark}
In the above specified formulas we assumed that the operator of causal evolution depends only on the time interval, but does not depend of the initial time moment: $U(t,t')= U_{t-t'}$. If the environment of the living being is varying with time, this assumption is invalid and one has to make use of the evolution operator $U(t,t')$ depending on two arguments. The formula (\ref{eq:corEvolution}) should then be appropriately modified. 
\end{remark}
\begin{remark}\label{remark:SysEnvir}
We assumed that the evolution of the environment is specified independently of the state of the living system. This may be justified in many cases. However, this assumption has to be abandoned in case of those criteria for postcorrection which include parameters of the environment as well as the parameters of the living system itself (such criteria will be considered in Sect.~\ref{sec:Criteria}). Then $\calH$ has to be defined as the space of states of the compound system including the living system and its environment. The operator $U(t,t')$ is then the evolution operator in this more wide space. 
\end{remark}
\begin{remark}
The evolution represented by the operator (\ref{eq:corEvolution}) consists of the series of operations, each being the causal evolution preceded by postcorrection. Such an evolution is characterized by two time parameters: \emph{the period of correction} $\tau$ and \emph{the depth of postcorrection} $T$. It is possible that some processes in living organisms are adequately presented by such a type of evolution (for example, higher animals and humans periodically experience the state of sleep in which the correction of the state of the organism is performed). However, continuous regime is typical for other correcting processes. In these cases an evolution law with continuous postcorrection should be applied. The simplest variant of it can be obtained as a limit of the discrete process. 
\end{remark}
\begin{remark}
We considered a transparent mathematical model of life in which the postcorrection is presented by a projector. This may be (and in fact should be) generalized. For example, the criterion for postcorrection may be presented by a positive operator (not a projector). This is evidently necessary for those criteria for postcorrection that are connected not with survival, but with less critical parameters of quality of life. Such criteria will be considered in Sect.~\ref{sec:Criteria}. 
\end{remark}
%%%%%%%%%%%%%%%%%%%%%%%%%%%%%%%%%%%%%%%%%%%%%%%%%%%%%%%%%%%%%%
%%%%%%%%%%%%%%%%%%%%%%%%%%%%%%%%%%%%%%%%%%%%%%%%%%%%%%%%%%%%%%

Up to now we considered only the simplest scheme for support of life of a single living being. This scheme requires only a single criterion of life called survival and mathematically presented by the projector $L$. This may be enough for primitive forms of life in the condition of unlimited resources (first of all food). However, for realistic description of more sophisticated forms of life one has to consider more complicated criteria. Besides, the role played by the living beings in respect to each other should be taken into account. 

All this requires further generalizations of the mathematical model of life. Not pretending to be quite general and precise in detail, we shall illustrate possibilities of such generalizations in some typical situations. In Sect.~\ref{sec:CollectiveLife} a sort of collective criterion of survival will be considered, and in Sect.~\ref{sec:Criteria} the classification of various criteria of life and corresponding aspects of the phenomenon of life will be presented. 

\begin{remark}
``A future state'' of a system has been used by Y.~Aharonov, P.G.~Bergmann and J.L.~Lebowitz in the paper published in 1964 \cite{ABL} and by Y.~Aharonov with other coauthors in the subsequent works (see for example \cite{AharonovVaidman,AharonovGruss}) under name of \emph{the formalism of postselection} or \emph{the two-vector formalism}. In this formalism the states of a system at both initial time and some later moment of time (``final time'') are fixed. In \cite{ABL} the formula for the probabilities of various outputs of the measurement performed at an intermediate time (between the initial and final times), given the initial and final states, was derived. The above defined operation of \emph{postcorrection} differs from the two-vector formalism (postselection) both formally and essentially. The formal difference is that in the postcorrection 1)~not a single state but a subspace (of an arbitrary dimension) is fixed in the future (at the ``final time''), and 2)~the initial state undergoes a correction. The essential difference is in the physical interpretation (sphere of application) suggested for these two formalisms. The two-vector formalism was applied for analyzing events predicted by conventional quantum mechanics for usual material systems. In the paper \cite{AharonovGruss} the two-vector formalism was exploited to formulate a novel interpretation of quantum mechanics, in which the various outputs of a measurement were associated with various future state vectors. In contrast with this, the postcorrection describes (in the framework of EEC) not a usual material system, but a ``living system'', or, more precisely, the image appearing in the consciousness of living beings.\footnote{In case of a primitive living being, the expression ``the image appearing in the consciousness'' stands for the information which is exploited by this living being to manage its behavior.}
\end{remark}

\newpage
\section{Collective criterion of survival}
\label{sec:CollectiveLife}

It was shown in Sect.~\ref{sec:Postcorrection} how evolution of a living being providing its survival may be described mathematically in terms of the operation of postcorrection. The simplest form of this operation considered in Sect.~\ref{sec:Postcorrection} was determined by a single criterion of survival which in turn was presented by a projector $L$ on the subspace of states in which the living system remains alive. This model is sufficient for simple forms of life and unlimited resources (first of all unlimited amount of food). 

Let us consider now the model of life in which resources are limited so that only a limited number of living beings can survive. 

It is clear that in this case the relations between various living beings become important and should be taken into account. One possible strategy for survival of living beings in these hard conditions is competition (fight) of them with each other. However, \emph{the collective strategy of survival} is also possible in this case. Let us consider the simplest mathematical model of such a collective strategy. 

Consider a group consisting of $N$ similar living beings (living systems), enumerated by the index $i\in \Omega$, where $\Omega = \left\{1,2,\dots,N\right\}$. The living system having the number $i$ is described by the state space $\calH_i$ and projector $L_i$ in this space as a criterion of survival. The corresponding orthogonal projector is $D_i$. The sum $L_i+D_i$ is a unit operator in the space $\calH_i$. The operators $L_i$ and $L_{i'}$ commute with each other because they act in different spaces $\calH_i$ and $\calH_{i'}$. Denote by $|I|$ the number of elements in the set $I$ and by $\bar{I} = \Omega\setminus I$ the complementary subset in $\Omega$ (the set of elements of $\Omega$ which are not elements of $I$). 

In the conditions of unlimited resources all living systems forming the group can exist (survive) independently of each other. Then each of them may be described by the simple model considered in Sect.~\ref{sec:Postcorrection} so that all of them can survive forever.\footnote{in the framework of the present simple model} Assume however that the resources (for example food) that can be found in the environment are limited and their amount is sufficient only for survival of $n$ living systems of this type. In this situation life may be regulated in such a way that the interests of the whole group are taken into account. Then a sort of a ``super-organism'' exists. This means that the group consisting of $N$ living beings behaves as a single living system. What has to be taken as a criterion of survival of the whole group in this case? 

The simplest form of the collective criterion of survival is following: 
$$
L_{(n)} = \sum_{I\subset \Omega,\; |I|=n} L_ID_{\bar I}
$$
where it is denoted $L_I=\prod_{i\in I}L_{i}$ and $D_{\overline{I}}=\prod_{i'\in \overline{I}}D_{i'}$. It is not difficult to show that this operator is a projector, and the projectors $L_{(n)}$ and $L_{(n')}$ are orthogonal for $n\neq n'$. The set of projectors $\left\{L_{(n)}|n=0,1,2,\dots,N\right\}$ form a complete system of orthogonal projectors.  

The correction described by the operator of survival $L_{(n)}$ guaranties that in the time interval $T$ precisely $n$ living systems will be alive, the rest will be dead. This means that the resources will be sufficient for those which are alive. The death of some members of the group is in this case a condition for survival of the rest.  

It is interesting in such a model that the correction of the state of the group of the living systems which is expressed by the operator $L_{(n)}$, describes not fighting the members of the group between each other, but rather collective regulation of their states. This regulation provides survival of the group with the maximal possible number of members. The state of each living system in the group is corrected at the present time moment, and thus corrected states, simply because of the natural evolution (described by the unitary operator $U_T$), results in the death of certain number of the members of the group. The number of those who have to die, is sufficient for surviving the rest in the conditions of the available resources. 

Such a correction of the state may be called \emph{collective programming of death} for some members of the group \emph{for the sake of life} of the rest. The collective program of death does not determine which members of the group have to die (the choice varies for various alternatives). Therefore, this is actually \emph{the strategy of collective survival} discriminating none. 

The well-known program leading to death of an organism in a certain age is a sort of the collective strategy of survival for the given species. In this case the reason for programming death is not the deficit of resources, but the task of the progress of the species as a whole. 

Evidently, in most groups of animals the survival is regulated by collective criteria. This explains particularly why intraspecific competition is as a rule absent. In this respect humans radically differ. It seems that they have collective criteria for the collectives (groups) of various levels: for a nation, for a social group, for a family and so on, up to the individual criteria. This makes possible conflicts between different groups of people. In the limit this may result in fighting anyone against anyone. 

In our time the exponential development of technology makes it available for small collectives. In these conditions individual criteria of survival and even lower levels of the collective criteria of survival (i.e. individualistic consciousness) increase violence so strongly that the very existence of Mankind is in danger. This crisis may be prevented only by the transition to the universal (common for all people and even for all living beings) collective criterion of survival (i.e. to collective consciousness). 

It was suggested long time ago \cite{TeyarEng,SatpremEng,Grof1997eng} that transition to the collective consciousness is necessary for preventing the global crisis. However, it is unclear up to now how the transition of most people to the collective consciousness may be achieved in practice (the catastrophe may be prevented only in case of most people changing their consciousness). The theory of consciousness following from EEC gives grounds for optimism. According to EEC, the change of the consciousness will happen automatically, the crisis will be stopped, and the catastrophe prevented.\footnote{The transition of almost all people to the universal criterion of survival and collective consciousness will necessarily happen in one of the alternatives at the moment of the highest level of the global crisis. The catastrophe will be therefore prevented in this alternative. Those people who have changed properly their consciousness beforehand, will witness just this alternative with high probability. Those who have not managed to change their consciousness, with high probability will watch the end of world.} 

\newpage
\section{Various criteria for postcorrection}
\label{sec:Criteria}

In the preceding sections we considered postcorrection with the criterion of survival, the most important criterion for living beings. In fact this criterion defines life as such. The simplest model exploiting only this criterion is sufficient to represent the simplest forms of life. However, other operations of postcorrection, based on other criteria of life, become actual for more sophisticated forms of life. The set of all criteria of life characterize quality of life in more detail. Several various operations of postcorrection, corresponding to various criteria, are performed in this case simultaneously. It seems plausible that in case of human beings criteria for postcorrection may exist which are connected not only with the parameters of the human organisms (bodies), but also with the parameters of their environment. 

Analyzing various criteria for postcorrection is an interesting problem that may be approached from various viewpoints. Not pretending to be general and precise in details, we can suggest a rough classification of possible criteria of life as follows. 
\begin{itemize}
	\item Criteria of survival
	\begin{itemize}
		\item The criterion of survival for a single creature
		\item The criterion of survival for a group of creatures
		\item The criterion of survival for the living matter as a whole
	\end{itemize}
	\item Parameters of the state of the body
	\begin{itemize}
		\item Evidence of being alive or dead (the criterion of survival)
		\item Various levels of the quality of life
		\item Immaterial parameters (insignificant for the quality of life)
	\end{itemize}
	\item Parameters of the environment (conditions for life)
	\begin{itemize}
		\item Parameters, which are essential for surviving
		\item Parameters, which are essential for the quality of life
		\item Immaterial parameters (insignificant for the quality of life)
	\end{itemize}
\end{itemize}

Let us make some remarks concerning this (of course, oversimplified and approximate) scheme of classification. 

It is clear that a sophisticated structure of living systems allows them to control not only survival, but also quality of life. In our mathematical model this may be described by the same scheme of postcorrection as in Sect.~\ref{sec:Postcorrection} but with projecting on a narrower space of states in which not only life keeps on but the quality of life remains sufficiently high. This suggests that in an arbitrary state from the given subspace \emph{the parameters of the state of the body} are in the limits characterizing the given quality of life. 

The question naturally arises why we included \emph{immaterial parameters} (those which are insignificant for the quality of life) in the list of the criteria for postcorrection. Without a doubt, the control on these parameters is unnecessary to provide the main internal needs of life. However, anyone knows from his own experience that at least human beings (but most probably also animals) are in command of certain immaterial parameters of their bodies and do control them. This reveals itself in the phenomenon of \emph{free will}. 

Indeed, a person can, according to his will, choose one or another variant of behavior with no essential influence on the fact of survival, or even on the quality of life. For example, he may in certain limits vary the schedule of his meals, amount of food he eats and its choice (the menu). The more so, one may decide quite arbitrarily whether he wish to open or close the window, to read a book or watch TV and so on. 

In the framework of our model a free will is an arbitrary choice of some immaterial parameters of the body, and execution of the free will is the postcorrection for a short time interval, performed according to the chosen criteria. 

\begin{figure}[t]
	\centering
		\includegraphics[width=0.40\textwidth]{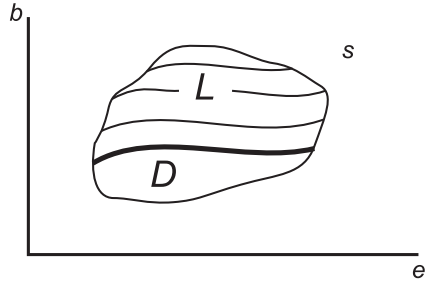}
	\caption{Various criteria for postcorrection: the state of the world $s$ is determined by the state of the body $b$ and the state of its environment $e$. The regions $L$ and $D$ correspond to survival and death. Horizontal lines separate the regions corresponding to different levels of the quality of life. Any subregion on the plane determines certain criterion according to which postcorrection may in principle be performed.}
	\label{fig:Fig1}
\end{figure}

Considering various parameters for postcorrection from somewhat different point of view, one may suggest the following (of course, also tentative) classification (see Fig.~\ref{fig:Fig1}). 
Denote by $s$ (after the word ``states'') the set of various parameters of life (characterizing both the body and the environment). The parameter $s$ is in fact a pair $s = (e,b)$, where $e$ (after the word ``environment'') stands for the conditions of life, or the state of the environment, and corresponds to the horizontal axis, while the parameter $b$ (after the word ``body'') refers to the state of the body of the living being (the bodies of a group of the living beings) and corresponds to the vertical axis. The parameter $s$ lies in some two-dimensional area, in which the very notion of life makes sense.\footnote{In reality each of the parameters $e$ and $b$ is multidimensional, thus we talk of the ``two-dimensional'' area only for the sake of an obvious image.} This area is divided with a horizontal line in two parts. The parameters in the upper part of the area correspond to survival (projector $L$), while the lower part corresponds to death (projector $D$). The region of survival is partitioned in the subregions corresponding to various levels of the quality of life.

Each subregion in the upper part of the area drawn in Fig.~\ref{fig:Fig1} determines some criterion according to which the postcorrection may in principle be performed (but is not necessarily performed in reality). Of course, in general case the criterion for postcorrection is defined as an operator in the space of states of the whole world rather than the states of the living system itself (as in the examples discussed in Sects.~\ref{sec:Postcorrection},~\ref{sec:CollectiveLife}). This is the situation when evolution of the compound system including both living system and its environment has to be considered (see Remark~\ref{remark:SysEnvir}). 

The operations of postcorrection with various criteria describe various aspects of the phenomenon of life. This may be illustrated by the following scheme of identifications. 
\begin{itemize}
	\item \emph{Life} (the principle of life, without details) = postcorrection with the criterion of survival for the living matter as a whole. 
	\item \emph{Survival} = postcorrection with the criterion of survival relating to the body (bodies). 
	\item \emph{Support of the health} = postcorrection with the criterion of quality of life relating to the body.
	\item \emph{Free will} = postcorrection with the criterion, relating to the own body, but as a rule immaterial for survival.\footnote{The exclusions such as suicide require more detailed model accounting for the influence of the living system onto its environment.} 
	\item \emph{Control on the appearing reality (probabilistic miracle)} = postcorrection with the criterion relating to an object outside the own body. 
\end{itemize}

The last point concerns an unusual phenomenon, called \emph{the probabilistic miracle}. By this term we mean that a human person, by the power of his consciousness, makes happen such an event in his environment which has low, though nonzero, probability (we suggest that some persons can do such things). The ability to perform probabilistic miracles does not seem to be necessary, in the usual meaning of the word, for life. However, first, this phenomenon naturally enters the general scheme so that its exclusion could look artificial, and, secondly, the human experience seems to point out that the events of this type really take place. 

There is one more class of unusual phenomena in the sphere of consciousness (and therefore in the sphere of life) that can be explained by postcorrection: 
\begin{itemize}
	\item \emph{Insight} = postcorrection with the criterion of truth
\end{itemize}
This class includes foresights, insights (among them scientific insights), \emph{direct sighting of truth} (i.e. conclusions not supported by logic or facts). All these phenomena can be explained in the following way.   

Let a person formulate some question or pose some problem (a scientific problem is a good example). Then, in order to experience insight, he has to go over to the regime of unconscious (not necessarily completely disabling his explicit consciousness but at least disconnecting it from the given problem). In this regime a faithful solution of the problem comes out sooner or later without any further effort, as an insight.\footnote{This does not mean that hard problems may be solved without any work. In order for the process to be efficient, the problem should be formulated and preliminarily worked out in much detail that requires hard work on the first stage.} 

In fact, the true solution of the problem is selected, with the help of the postcorrection, among all thinkable ``attempted solutions'', most of them wrong. 

The selection is performed in this case with the help of \emph{the postcorrection with the criterion of truth}. Even if the problem cannot be solved at the present time by conventional methods (on the basis of the known facts and logical conclusions), it may have evident solution in future. For example, some future events may point to the correct solution. In case of a scientific problem new experiments may be realized in future which unambiguously point to the right solution, singling it out from all seemingly possible ``attempted solutions'' of the problem. Therefore, a criterion of the true solution of the given problem may exist in the future even if it is absent in the present. 

In all these cases the operation of postcorrection does correct the present state making it to be in accord to the criterion existing in the future. This results in the immediate choice of the correct solution of the problem, although its correctness can be confirmed only in the future. Consciousness, when being in the regime of unconscious, obtains the ability to look into the future, and makes use of the obtained information in the present. 

The idea may be clarified if it is reformulated in terms of the states of brain. From all states of the brain corresponding to various ``attempted solutions'' of the problem (wrong ideas of the solution among them) the postcorrection selects the state which corresponds to that solution which has to be confirmed in the future. This change of the state of the brain means that insight, or direct sighting of truth, occurred. 

By the way, it is known from the experience that the person applying this process for solving a problem, feels to be absolutely confident that the solution guessed by him in the course of the insight is true. This is not at all strange because the solution found in this way is not a product of his imagination but the genuine true observed by the mechanism of direct sighting. 

Great scientists, Albert Einstein among them, confirm the fact that they always feel absolute confidence in the solution found in the insight, and the solution found in this way always turns out to be correct in the course of its verification by conventional methods. 

An interesting remark may be made about the criterion of truth used in the process thus described. This criterion may sometimes coincide with the ``formal proof'' which is found by the scientist after he had experienced instantaneous insight. It is clear that the formal proof may serve as a criterion of truth for a solution of the given problem. This criterion does not exists (not yet found) at the moment of the insight, but it arises later on, when the solution having been guessed in the insight is later deduced by conventional methods. The whole process looks like lifting oneself by hairs. Does it really supply any advantage for solving the problem? Let us show it does. 

Solving any problem is easier if it is known that the solution exists (may be it is known that this problem has already been solved by someone else) and much easier if the final result (not its proof) is available. Just this situation of the final solution known beforehand is realized in the process of the scientific insight followed by the formal derivation of the foreseen solution. Indeed, the scientist anticipates the right solution in the course of insight, he is completely confident in this solution, and because of this it becomes much easier for him to formally derive the foreseen solution by conventional methods. It is curious that in this case the scientist foresees the certainly right answer which himself will find in some time.\footnote{This ability is very exciting in case of great scientists, but it is often is exploited by many experienced scientists as well as people from other professions and simple people in the each-day life.} 

The operator of postcorrection selecting the right solution of the problem (among ``the attempted solutions'') may be presented in the form $P_T = U_T^{-1}PU_T$, where $P$ is a criterion of the correct solution. The operation of postcorrection presented by the operator $P_T$ is efficient if the criterion $P$ is not realizable at present, but can be realized in the time interval $T$. 

This leads us to the question about the role of brain. Many attempts to explain how work of brain can produce the phenomenon of consciousness gave in fact no result. In each of these attempts either a logical circle is included (what should be proved is implicitly assumed) or not consciousness as such is dealt with in the argument, but various operations performed in the consciousness (for example, calculations or logical conclusions). 

From the point of view of the theory we consider here, EEC, consciousness is not a product of brain, but a separate, independent phenomenon, closely connected with the very concept of life. What about brain, it is an instrument of consciousness rather than its origin. 

The brain is used by the consciousness to control the body and obtain information about its state (and, through its perception, about the state of the environment). In other words, the brain (or rather some regions in it) is the part of the body which realizes its contact with the consciousness, it is an interface between the consciousness and the body as a whole. In particular, when it is necessary the brain forms the queries that should be answered. Sometimes these queries are answered by the brain itself with the help of the processes of the type of calculations and logical operations. Other queries cannot be solved directly in the brain and are solved by the consciousness with the help of ``direct sighting of truth'' (by postcorrection). 

\begin{remark}
A. Losev and I. Novikov noted \cite{NovikovEng} that time machines (space-times including closed timelike curves), in case if they exist, may be used for solving mathematical problems with the help of the methods or technical devices which are not known at present but can be realized in future. For this aim, the problem is solved at the time when the necessary methods are created and then its solution is sent into the past with the help of the time machine. The above formulated mechanism for solving problems (of arbitrary types) with the help of postcorrection is quite analogous. The only difference is that the ``time machine'' acting in this process is virtual and ``exists'' only in human consciousness. 
\end{remark}

\newpage
\section{Conclusion}
\label{sec:Conclusion}

Extended Everett's Concept (EEC) originated as an attempt to improve the interpretation of quantum mechanics proposed by Everett. Nevertheless, it is not simply a novel interpretation, but in fact a theory going beyond the framework of quantum mechanics. Starting from the role played by consciousness in the conceptual problems of quantum mechanics, EEC finally results in understanding what is consciousness and, more generally, what are specific features of living matter. 

Considering consciousness on the basis of EEC, one is led to the conclusion that the conventional (causal) laws of nature are insufficient for describing phenomenon of life. The laws of nature elaborated in physics (including quantum physics), chemistry and other natural sciences correctly describe the behavior of non-living matter. The behavior of ``living matter'' cannot be explained only by the action of usual laws of nature (say, quantum mechanics). Nevertheless, comprehensive analysis of quantum mechanics indicates at the principal points in which the laws acting in the sphere of life have to differ from the conventional physical laws. The laws governing living matter may then be formulated at least in their most general aspects. Just this is made in EEC. 

The novel features that have to be introduced in order to describe the phenomenon of life, can be formulated in various ways. Restricting himself by the most general formulation, one may say that not only causes but also goals play role in behavior of living matter. The main goal, always existing in connection with living beings, is survival, or persistence of life (this may be survival of a single living being, or of some group, for example of a herd or specie of animals). Therefore, the goal of survival has to be accounted in the evolution law for living matter. 

In the preceding works of the author on EEC \cite{EEC2000eng,EEC2005eng,EEC2005bkEng,EEC5} the laws governing life were formulated on the basis of the concept of consciousness and its identification with the separation of alternative classical realities (the concept characteristic of the Everett's interpretation). In this context the term ``consciousness'' embraces not only the explicit consciousness, but also the sphere of unconscious. Moreover, just in the regime of unconscious (or at the border between the explicit consciousness and unconscious) those features of consciousness are revealed which is the very essence of the phenomenon of life: the ability to obtain information from all alternative realities and select those alternatives which are most favorable for life. 

In the present paper we have shown that the evolution of ``a living system'' (following from EEC) can be described mathematically if one introduce, besides the usual (unitary) evolution operator, an additional operation called postcorrection. This operation corrects the state of a ``living system'' to provide necessary features of this state in future: survival of the living system or even certain quality of its life (for example the health). Introducing postcorrection in the evolution law of the living system allows one to classify various forms of life and various aspects of the phenomenon of life, depending on what characteristics of life can be provided by the postcorrection. We shortly discussed only the key points of this classification. The detailed elaboration of the theory is a question of its future development. 

The operation of postcorrection not only supplies a mathematical formulation of the principal feature of EEC, but also simplifies the logical structure of this theory. In fact, \emph{it is sufficient to postulate that the boundaries of the sphere of life are governed by postcorrection}. After this, the concretization of the theory requires only the choice of the criteria, according to which the postcorrection is performed. 

Unexpected (from the physical viewpoint) interpretation of the operation of postcorrection, as describing evolution of living matter, became possible because we did not restrict ourselves strictly by the framework of physics. Starting from the arguments originated in physics (conceptual problems of quantum mechanics) and following the ideas of EEC, we were forced to go beyond the limits of physics as such and to consider at least the principal points of theory of living matter. Instead of the known (accepted in physics) statement that each event has its own cause, we had to agree that all important events and processes in the sphere of life are determined not only by causes but also by goals, first of all by the goal of survival. In the resulting theory the operation of postcorrection is a mathematical formalization of the almost evident fact that the goals play central role in evolution of living matter. 

Let us remark that theory of consciousness and life following from EEC essentially differs from the usual mechanistic approach which considers the phenomenon of consciousness as a function of brain. From the viewpoint of theory of ``quantum consciousness'' resulting from EEC, the brain is rather an instrument exploited by the consciousness (as a specific feature of a ``living system'') to control the body and obtain information about the environment through the body and its organs. 

This, by the way, allows one to look in another way at the problem of artificial intellect. The conclusion following from EEC is that it is possible to create an automat possessing intellectual abilities (there are great achievements in this respect nowadays), but it is principally impossible to create a machine having consciousness as something that can to perform postcorrection, i.e. such that can be called ``artificial living being''. 

The postulate of postcorrection broadens quantum mechanics, including in the consideration the law of evolution of living matter. The resulting theory is in a way symmetrical in time direction. Non-living matter evolves in the causal way (the past determines the future), but in the sphere of life only those initial conditions are left which provide survival (the future determines the past). This ``influence of the future on the past'' is realized as the selection of favorable scenarios and mathematically described by postcorrection. 

Let us make finally one more remark demonstrating how natural for living systems the evolution law (\ref{eq:corEvolution}) including postcorrection is. This law is, in its spirit, very similar to what is called \emph{the antropic principle}. The antropic principle explains the special ``fine tuning'' of the parameters of our world by the fact that in case of any other set of the parameters organic life would not be feasible and therefore no humans could exist to observe this world. The principle of life, formulated as the ability of the living system to postcorrect its state and provide its survival, suggests in fact something quite similar, even in a softer variant. 

In order to explain this, we have to underline once more that \emph{the postcorrection describes selecting those scenarios which have to remain in the sphere of life}. The rest scenarios do not disappear. They are just as real as those selected, but they are not included in the sphere of life, i.e. an observer cannot watch these ``unfavorable for life'' scenarios. \emph{``The sphere of life'' is such an image of our world which can be observed}. If just this image (i.e. not ``the whole world'' but only ``the sphere of life'') is taken as a starting point for constructing evolution law, then the result of the construction will necessarily be the evolution including the postcorrection.

Thus, postcorrection in the evolution of the living matter (of the sphere of life) does not need even being postulated. Instead of this it may be derived from the (generalized) antropic principle. Non-living matter satisfies the usual quantum-mechanical evolution law. The evolution of the living matter (of the sphere of life) simply by definition should include postcorrection.

\end{document}